\documentstyle[12pt]{article}

\newcommand{\be}{\begin{equation}}
\newcommand{\ee}{\end{equation}}

\begin{document}
\pagestyle{empty}
\title{Nambu--Jona-Lasinio Models\\ Beyond the Mean-Field
Approximation}
\author{P. P. Domitrovich, D. B\"uckers, and H.\ M\"{u}ther
\\\\
Institut f\"{u}r Theoretische Physik,\\ Universit\"{a}t T\"{u}bingen,\\
D-7400 T\"{u}bingen, Germany}
\maketitle

\date{\today}

\begin{abstract}
Inspired by the model of Nambu and Jona-Lasinio, various Lagrangians are
considered for a system of interacting quarks. Employing standard
techniques of many-body theory, the scalar part of the quark self-energy
is calculated including terms up to second-order in the interaction.
Results obtained for the single-particle Green's function are compared
with those which only account for the mean-field or
Hartree-Fock term in the self-energy. Depending on the explicit form of
the Lagrangian, the second-order contributions range between 4 and 90
percent of the leading Hartree-Fock term. This leads to a considerable
momentum dependence of the self-energy and the effective mass of the quarks.
\end{abstract}
\pagestyle{empty}


\clearpage
\pagestyle{headings}

\section{Introduction}

The so-called NJL model of Nambu and Jona-Lasinio was proposed
more than 30 years ago\cite{njl}. In its original form, this model was
constructed as a theory of nucleons interacting via an effective
two-body interaction. Today, the Fermions described by this model are
usually reinterpreted as quarks, and, with this interpretation, the NJL model
has become extremely popular over the past few years (see, e.g., the
recent review by Klevansky\cite{klev} and the references quoted within it).

There are quite a few reasons for the popularity of the NJL model: A
very important one is the fact that the Lagrange density is constructed
to obey some of the symmetries of QCD. In particular, a model is constructed
which permits the investigation of chiral symmetry of light quarks and
the dynamical breaking of this symmetry. Dynamically broken chiral symmetry
generates effective masses for the constituent quarks which are much
larger than the bare current quark masses.

The attraction of the NJL model has increased as a result of successes
regarding systematic attempts to
determine effective theories which account for the most relevant
degrees of freedom of QCD at low energies. It has been shown that the
NJL model can be understood as a low-energy approximation to QCD
\cite{schad,alkof}. A bosonization of the NJL Lagrangian leads to an
effective Lagrangian in terms of meson fields, which in general shows
satisfactory agreement with low-energy meson data \cite{ebert}. In this
sense, the NJL model can be described as an effective quark theory based on
low-energy QCD, which permits the evaluation of certain observables.
For example, it may provide useful predictions concerning properties such as
the medium- or temperature- dependence of masses and coupling constants of
low-energy hadrons\cite{ulf,hats}, despite the fact that it does not predict
confined quarks.

Most studies of the NJL model in its Fermionic
representation have been restricted to the mean-field or Hartree-Fock
(HF) approximation. In the present investigation, we advance beyond
the HF approach and include in the definition of the irreducible
self-energy of the quarks all terms up to second-order in the
interaction (see figures 1a) and 1b)). The second-order terms of figure
1b), describing the coupling of a quark to a configuration of a quark $q$
plus a $q\bar q$ excitation, can be understood as the term of lowest-order
describing the contribution to the quark self-energy by the
exchange of a collective $q\bar q$ excitation, i.e.,
meson-exchange. Note that, however, in contrast to, e.g., Cao et.al.
\cite{cao}, we not only want to consider the effects of $\pi$-exchange, but
also to include all relevant $q\bar q$ channels accounting for collective
as well as non-collective modes.

Having determined the quark self-energy in this improved approximation,
one could determine the corresponding single-particle Green's function and
evaluate the polarization propagator, which contains the information about
the $q\bar q$ excitation modes. This polarization
propagator contains the contributions represented by the diagrams of
figures 1c) - 1e), among others, where the Fermion lines represent
the Green's functions
of the bare quarks. The diagram of figure 1e), in particular, is taken
into account by our improved definition of the quark self-energy. This
contribution might be significant, especially for the polarization
propagator in the scalar-isoscalar channel describing the so-called
$\sigma$ meson. In this case, the coupling of the $q\bar q$ excitation
modes to the $2q2\bar q$ configurations, like, e.g., the $2\pi$ states should
be important. It is well known that the simple NJL model predicts a
$\sigma$ meson with a mass of about twice the constituent quark mass.
This scalar meson is frequently compared to the $\sigma$ meson, which is
required for a One-Boson-Exchange description of the nucleon-nucleon
interaction\cite{mach}. This meson is more representative of the
exchange of two correlated pions than a well defined $q\bar q$ meson.

However, evaluating the polarization propagators for the various meson
modes by employing single-particle Green's functions with the inclusion
of terms like diagram 1b) in the self-energy for quarks, but using the
bare interaction in the Bethe-Salpeter equation, would lead to an
approach which obviously does not fulfill the Goldstone theorem. This
means that, even for a Lagrangian which is truly invariant under a chiral
transformation, the mass of the Goldstone boson associated with this
symmetry would be different from zero. An approach which obeys the
Goldstone theorem can only be expected if the single-particle
propagator and the residual interaction used to calculate the
polarization propagator are chosen in a consistent way. This implies
that besides the diagram displayed in figure 1e), terms
represented by figure 1f) must also be taken into account. Diagram 1f)
can be interpreted as the leading contribution of the so-called induced
interaction, which has been investigated in the many-body theory of
Fermi liquids for such cases as nuclear matter \cite{rev,wim0} and
liquid $^3He$ \cite{babu}.

The present investigation should be considered as a first step towards
such a consistent treatment of the NJL model beyond the mean-field
approximation. For that purpose, we want to study the importance of the
second-order terms in the self-energy (figure 1b) as compared to the
Hartree-Fock contribution displayed in figure 1a). Several Lagrangians
will be considered, adjusting the coupling strength in such a way that
the same constituent mass is obtained for quarks of zero momentum in
each model. Following this introduction, section 2 will describe the various
Lagrangians employed, and section 3 sketches the techniques used to evaluate
the self-energy. The
results of the calculations are discussed in section 4, and the main
conclusion will be summarized in section 5.

\section{Various Lagrangians for a NJL model}

In its original form \cite{njl}, the NJL model was designed to describe
a system of interacting nucleons, but has more recently been reinterpreted
as a quark Lagrangian of identical form:
\be
{\cal L}_{D}^{(A)}={\bar\psi}(i\nabla\hskip-0.8em{/}-m_0)\psi
+G_{A}\left[(\bar\psi\psi)^2
+(\bar\psi{i}{\gamma}^{5}\vec\tau\psi)^2\right], \label{eq:njla1}
\ee
such that the Fermion field $\psi$ now represents
quarks with SU(2) flavor ($\vec\tau$ denoting the Pauli matrices for
the flavor degrees of freedom) and SU(3) color.
For vanishing current
quark mass $m_{0}$, this Lagrangian is invariant under a chiral
transformation.
Particular to this study, A denotes the first of the four models considered
and D implies that only the direct part of the contact interaction defined in
eq.(\ref{eq:njla1}) is included.

Due to the presence of a contact interaction, the NJL model is
nonrenormalizable and therefore requires a regularization that can be
effected through the use of a cut-off scheme.
Throughout this study, we will use the
simplest, three-momentum noncovariant cut-off scheme. This means that
all integrals over internal momenta are restricted to three-momenta
${\bf p}^2$ less than a cut-off parameter $\Lambda^2$ after carrying
out the $p_{0}$ integration. As most of the qualitative features
of the model are not
very sensitive to the kind of regularization scheme employed, we expect
that the qualitative results obtained in the present study should also be
independent of the cut-off procedure.

Note that the quark-quark interaction in the model defined by
eq.(\ref{eq:njla1}) is assumed to be a pure color-scalar. Therefore, we
have suppressed the unit SU(3) operator acting on the color part of the
spinors $\psi$. The color degrees of freedom become important when
we want to account for the effects of the exchange terms in the two-Fermion
interaction. For a contact interaction like the one defined in
eq.(\ref{eq:njla1}), the exchange terms can easily be determined by
writing down the Fierz transformation of the direct interaction part in
eq.(\ref{eq:njla1}) and subtracting the exchange terms generated in
this way from ${\cal L}_{D}$:
\begin{eqnarray}
{\cal L}^{(A)}& = &{\cal L}_{D}^{(A)} - {\cal L}_{E}^{(A)} \\
&= & {\bar\psi}(i\nabla\hskip-0.8em{/}-m_0)\psi
+\frac{13}{12}G_{A}\left[(\bar\psi\psi)^2
+(\bar\psi{i}{\gamma}_{5}\vec\tau\psi)^2\right]  -
\frac{1}{4}G_{A}(\bar\psi{\gamma}_{\mu}\vec\lambda\psi)^2
\nonumber \\
&&+\frac{1}{8}G_{A}\left[(\bar\psi i\gamma_{5}\vec{\tau}\vec{\lambda}\psi)^2
+(\bar\psi\vec\lambda\psi)^2\right]
-\frac{1}{8}G_{A}\left[(\bar\psi\vec\tau\vec\lambda\psi)^2
+(\bar\psi i\gamma_{5}\vec\lambda\psi)^2\right] \nonumber \\
&& -\frac{1}{12} G_{A}\left[(\bar\psi i\gamma_{5}\psi)^2
+(\bar\psi\vec\tau\psi)^2\right] - \frac{1}{6} G_{A}(\bar\psi{\gamma}_{\mu}
\psi)^2 + \dots \label{eq:njla2}
\end{eqnarray}
One can see that the inclusion of the exchange terms not only leads to a
renormalization of the coupling constant for those terms already
existing in the direct part, but also generates additional interaction
terms of both the color-scalar and color-vector type ($\vec\lambda$ refers to
the Gell-Mann matrices acting on the color component of $\psi$). To
shorten the notation, we have omitted the tensor and pseudovector
components in the interaction. Also, the mesons associated with the tensor and
pseudovector interactions are not considered as relevant at nuclear length
scales and are therefore neglected in this study.

An alternate choice for a Lagrangian, which in the limit of massless
quarks is also symmetric under a chiral transformation and contains a
color-scalar direct interaction term, is given by:
\be
{\cal L}_{D}^{(B)}={\bar\psi}(i\nabla\hskip-0.8em{/}-m_0)\psi
-G_{B}(\bar\psi{\gamma}_{\mu}\psi)^2. \label{eq:njlb1}
\ee
Including the exchange terms in the same way as discussed above provides:
\begin{eqnarray}
{\cal L}^{(B)}&=&{\bar\psi}(i\nabla\hskip-0.8em{/}-m_0)\psi
-\frac{13}{12}G_{B}(\bar\psi{\gamma}_{\mu}\psi)^2
-\frac{1}{8}G_{B}\left[(\bar\psi\gamma_\mu\vec\lambda\psi)^2
+(\bar\psi{\gamma}_{\mu}\vec\tau\vec\lambda\psi)^2\right]
 \nonumber \\
&&-\frac{1}{12}G_{B}(\bar\psi{\gamma}_{\mu}\vec\tau\psi)^2
+\frac{1}{6}G_{B}\left[(\bar\psi\psi)^2
+(\bar\psi\vec\tau\psi)^2
+(\bar\psi{i}{\gamma}^{5}\psi)^2
+(\bar\psi{i}{\gamma}^{5}\vec\tau\psi)^2\right] \nonumber\\
&&+\frac{1}{4}G_{B}\left[(\bar\psi\vec\lambda\psi)^2
+(\bar\psi\vec\tau\vec\lambda\psi)^2
+(\bar\psi{i}{\gamma}^{5}\vec\lambda\psi)^2
+(\bar\psi{i}{\gamma}^{5}\vec\tau\vec\lambda\psi)^2\right] + \dots
\label{eq:njlb2}
\end{eqnarray}

Starting from the path integral formulation of QCD, attempts have been
made to eliminate the gluon degrees of freedom by trying to derive
or motivate effective Lagrangians for quarks in the low-energy domain.
Employing a current expansion of the effective quark action
\cite{roberts}, or using the field strength approach \cite{schad,hugo},
one obtains a quark-quark interaction defined in terms of color-vector
currents. Using the nomenclature introduced before, the
corresponding NJL type Lagrangian would be written as:
\be
{\cal L}_{D}^{(C)}={\bar\psi}(i\nabla\hskip-0.8em{/}-m_0)\psi
-G_C(\bar\psi{\gamma}_{\mu}\vec\lambda\psi)^2.
\label{eq:njlc1}
\ee
After the inclusion of the exchange terms, we obtain:
\begin{eqnarray}
{\cal L}^{(C)}&=&{\bar\psi}(i\nabla\hskip-0.8em{/}-m_0)\psi
-\frac{11}{12}G_C(\bar\psi{\gamma}_{\mu}\vec\lambda\psi)^2
-\frac{4}{9}G_{C}\left[(\bar\psi{\gamma}_{\mu}\psi)^2
+(\bar\psi{{\gamma}_{\mu}}\vec\tau\psi)^2\right] \nonumber \\
&&+\frac{1}{12}G_C(\bar\psi{\gamma}_{\mu}\vec\tau\vec\lambda\psi)^2
-\frac{1}{6}G_{C}\left[(\bar\psi\vec\lambda\psi)^2
+(\bar\psi\vec\tau\vec\lambda\psi)^2
+(\bar\psi{i}{\gamma}^{5}\vec\lambda\psi)^2
+(\bar\psi{i}{\gamma}^{5}\vec\tau\vec\lambda\psi)^2\right] \nonumber \\
&&+\frac{8}{9}G_{C}\left[(\bar\psi\psi)^2
+(\bar\psi\vec\tau\psi)^2
+(\bar\psi{i}{\gamma}^{5}\psi)^2
+(\bar\psi{i}{\gamma}^{5}\vec\tau\psi)^2\right] + \dots
\label{eq:njlc2}
\end{eqnarray}

Finally, we would also like to consider a Lagrangian of the form
that has been considered in ref. \cite{bueck} (without introducing
a momentum dependence of the coupling constant):
\be
{\cal L}_{D}^{(D)}={\bar\psi}(i\nabla\hskip-0.8em{/}-m_0)\psi
+G_{D}\left[(\bar\psi\vec\lambda\psi)^2
+(\bar\psi{i}{\gamma}^{5}\vec\tau\vec\lambda\psi)^2\right],
\label{eq:njld1}
\ee
which leads to:
\begin{eqnarray}
{\cal L}^{(D)}&=&{\bar\psi}(i\nabla\hskip-0.8em{/}-m_0)\psi
+\frac{11}{12}G_{D}\left[(\bar\psi\vec\lambda\psi)^2
+(\bar\psi{i}{\gamma}^{5}\vec\tau\vec\lambda\psi)^2\right]
+\frac{1}{6}G_{D}(\bar\psi{\gamma}_{\mu}\vec\lambda\psi)^2 \nonumber \\
&&-\frac{8}{9}G_{D}(\bar\psi{\gamma}_{\mu}\psi)^2
+\frac{4}{9}G_{D}\left[(\bar\psi\psi)^2
+(\bar\psi{i}{\gamma}^{5}\vec\tau\psi)^2\right] \nonumber \\
&&+\frac{1}{12}G_{D}\left[(\bar\psi{i}\gamma_5\vec\lambda\psi)^2
+(\bar\psi\vec\tau\vec\lambda\psi)^2\right]
-\frac{4}{9}G_{D}\left[(\bar\psi\vec\tau\psi)^2
+(\bar\psi{i}{\gamma}^{5}\psi)^2\right] + \dots \label{eq:njld2}
\end{eqnarray}

Comparing eqs.(\ref{eq:njla2}),(\ref{eq:njlb2}),(\ref{eq:njlc2}), and
(\ref{eq:njld2}), one can see that these four models yield quite
different ratios between the coupling constants of the corresponding
individual terms.
Assuming the same cut-off parameter $\Lambda$, one would obtain in the
Hartree-Fock approximation applied to the models (A-D) the same
constituent quark mass $m_{HF}^*$ from the so-called gap equation:
\be
m_{HF}^* = m_{0} + \tilde G\frac{24}{(2\pi )^3} \int_{0}^\Lambda d^3p \,
\frac{m_{HF}^*}{\sqrt{{\bf p}^2 + {m_{HF}^*}^2}} , \label{eq:gap}
\ee
if the coupling strengths in front of the scalar-isoscalar terms in
eqs.(\ref{eq:njla2}),(\ref{eq:njlb2}),(\ref{eq:njlc2}), and
(\ref{eq:njld2}) are chosen to be identical to $\tilde G$. This means:
\be
\tilde G = \frac{13}{12}G_{A} = \frac{1}{6}G_{B} = \frac{8}{9}G_{C} =
\frac{4}{9}G_{D} . \label{eq:coupco}
\ee
However, by choosing the coupling constants in this way, we see that the
strength in front of the other interaction terms will be quite
different. Therefore, depending on which model (A-D) we choose, we obtain
different strengths for the residual interaction in the various
channels even if we adjust the coupling constants in such a way that the
Hartree-Fock approximation leads to the very same gap equation.

\section{Calculation of the Self-Energy}

In calculating the self-energy contributions that are second-order in
the interaction, we will use the identification of momenta exhibited
in figure 2. Note that we need only consider diagrams representing
the direct terms since the exchange terms are taken into account by means of
a Fierz transformation, as discussed above. Adopting the Feynman rules
presented on pp. 107-108 of \cite{serot},
the second-order term can be written as:
\be
\Sigma_{\gamma\delta}(p) = i \sum_{\epsilon\eta} \int \frac{d^4q}{(2\pi
)^4} \Gamma_{\gamma\epsilon} (p-q) \Delta (p-q) g_{\epsilon\eta} (q)
\Gamma_{\eta\delta} (p-q) \Pi^{(0)}(p-q) .
\label{eq:self1}
\ee

As an example, we will discuss the case in which all interaction
vertices are given by the scalar part of the Lagrangian:
\be
{\cal L} = \tilde G (\bar\psi\psi)^2. \label{eq:scall}
\ee
This implies that the vertices $\Gamma$, which are independent of the
momentum variables, are given by:
\be
\Gamma_{\alpha\beta}(q) = \sqrt{2 \tilde G} \delta_{\alpha\beta} ,
\label{eq:verscal}
\ee
with $\alpha$ and $\beta$ referring to the indices of the color matrices,
flavor matrices, and Dirac spinors. In eq.(\ref{eq:self1}), $g_{\epsilon\eta}
(q)$ refers to the propagator of a quark with momentum $q$ and is multiplied
by an identity matrix in color space and flavor space.
Also, the ``interacting boson propagator'' $\Delta (p-q)$ can
be replaced by (-1), due to the contact interaction.
The polarization
propagator in its irreducible (lowest order form) $\Pi^{(0)}$ is
defined for the various excitation modes $\lambda$ by:
\be
\Pi^{(0)}_{\lambda} (q) = -i \int \frac{d^4k}{(2\pi )^4} \mbox{tr}
\left[ \Gamma_{\mu\nu}^\lambda (q) g_{\nu\rho}(k)
\Gamma_{\rho\sigma}^\lambda (q) g_{\sigma\mu}(k-q) \Delta(q) \right] ,
\ee
where the trace is over the three sets of indices mentioned above.
Inserting the vertices for the scalar excitation modes
(eq.(\ref{eq:verscal}), $\lambda = s$), and assuming for the Green's
function $g$ a Hartree-Fock self-energy for quarks which is characterized by a
constant effective mass $m^*$
(see eq.(\ref{eq:gap})), the imaginary part of the polarization propagator
at zero Fermi energy is given by:
\begin{eqnarray}
\mbox{Im} \Pi^{(0)}_{s} (q)& = &\frac{\tilde G n_{f} n_{c}}{4 \pi^2} \int
\frac{d^3k}{E^*_{k}E^*_{k-q}} \left( E^*_{k}E^*_{k-q} + {\bf k} ({\bf
k}- {\bf q}) - {m^*}^2 \right) \nonumber \\
&& \times \delta (\vert q_{0}\vert - E^*_{k} -
E^*_{k-q} ) \Theta (\Lambda - \vert {\bf k} \vert )
\Theta (\Lambda - \vert {\bf k } - {\bf q} \vert ) \; , \label{eq:pols}
\end{eqnarray}
where $n_{f}$ and $n_{c}$
stand for the number of flavors (2) and colors (3), respectively.
Furthermore,
\be
E^*_{k} = \sqrt{ {\bf k}^2 + {m^*}^2} .
\ee
The real part of the polarization propagator is related to the
imaginary part by a dispersion relation:
\be
\mbox{Re} \Pi^{(0)}_{\lambda} (q) =
\frac{1}{\pi} \mbox{P} \int_{0}^\infty dq_{0}' \,
\mbox{Im} \Pi^{(0)}_{\lambda}
(q_{0}', {\bf q} ) \left[ \frac{1}{q_{0}' - q_{0}}
+ \frac{1}{q_{0}' + q_{0}} \right] . \label{eq:disp}
\ee
In this manuscript,
we will focus our attention on the scalar part of the self-energy,
which is obtained from eq.(\ref{eq:self1}) with the aid of the relation:
\be
\Sigma^{(s)} = \frac{1}{4n_{f}n_{c}} \mbox{tr} \left( \Sigma \right) .
\ee
Taking the real part of the scalar part of the self-energy for the
scalar interaction and utilizing the dispersion relation
(\ref{eq:disp}), we obtain:
\begin{eqnarray}
\delta\mbox{Re}\Sigma_{s}^{(s)}
(p)&=&\mbox{Re}\left[ \frac{-i \tilde G}{2{n_f}{n_c}} \int
\frac{d^4q}{(2\pi )^4} \mbox{tr} \left[ g(q) \right] \Pi^{(0)}_{s} (p -
q ) \right] \label{eq:sigs1} \\
&&=-\frac{\tilde G m^*}{4 \pi^3}
\int_{0}^\Lambda \frac{{\bf q}^2  \, dq}{E^*_{q}} \int_{-1}^{+1} dx
\nonumber \\ &&
\times
\biggl\lbrace  \mbox{P} \int_{0}^\infty dq_{0} \frac{\mbox{Im}
\Pi^{(0)}_{s} (q_{0}, \vert {\bf p} - {\bf q} \vert )}{q_{0}+p_{0}+E_{q}^*}
+
\mbox{P} \int_{0}^\infty dq_{0} \frac{\mbox{Im}
\Pi^{(0)}_{s} (q_{0}, \vert {\bf p} - {\bf q} \vert )}{q_{0}-p_{0}+E_{q}^*}
\biggr\rbrace.
\nonumber \\ &&
\label{eq:sigs2}
\end{eqnarray}
The integration variable $x$ stands for the cosine of the angle between
${\bf p}$ and ${\bf q}$. From eq.(\ref{eq:sigs2}), one can verify
immediately that the scalar contribution indeed yields results which
only depend on the absolute value of the zero component $p_{0}$.

If, instead of the scalar part of the interaction (\ref{eq:scall}), one
now considers a pseudoscalar interaction:
\be
{\cal L} = \tilde G (\bar\psi i \gamma^5\psi)^2, \label{eq:pseudl}
\ee
the corresponding imaginary part of the polarization propagator is:
\begin{eqnarray}
\mbox{Im} \Pi^{(0)}_{ps} (q)& = &\frac{\tilde G n_{f} n_{c}}{4 \pi^2} \int
\frac{d^3k}{E^*_{k}E^*_{k-q}} \left( E^*_{k}E^*_{k-q} + {\bf k} ({\bf
k}- {\bf q}) + {m^*}^2 \right) \nonumber \\
&& \times \delta (\vert q_{0}\vert - E^*_{k} -
E^*_{k-q} ) \Theta (\Lambda - \vert {\bf k} \vert )
\Theta (\Lambda - \vert {\bf k } - {\bf q} \vert ) \; .
\label{eq:polps}
\end{eqnarray}
Eq.(\ref{eq:polps}) can be used to evaluate the contribution to
the scalar part of the self-energy originating from eq.(\ref{eq:pseudl}):
\be
\delta\mbox{Re}\Sigma_{s}^{(ps)}
(p)=\mbox{Re}\left[ \frac{-i \tilde G}{2{n_f}{n_c}} \int
\frac{d^4q}{(2\pi )^4} \mbox{tr} \left[ i\gamma^5 g(q) i\gamma^5 \right]
\Pi^{(0)}_{ps} (p -q) \right] . \label{eq:pses1}
\ee
This yields a result similar to eq.(\ref{eq:sigs2}), except that one
must replace $\Pi^{(0)}_{s}$ by $\Pi^{(0)}_{ps}$ and supply an
overall minus sign, since
$\mbox{tr} [ i\gamma^5 g(q) i\gamma^5 ] = -\mbox{tr} [ g(q) ]$.

Finally, we consider
the contribution of second-order originating from a vector interaction:
\be
{\cal L} = -\tilde G (\bar\psi \gamma_{\mu}\psi)^2 . \label{eq:vecl}
\ee
In this case, the contribution to the self-energy is given by:
\begin{eqnarray}
\delta\mbox{Re}\Sigma_{s}^{(v)} (p)
&=& \frac{\tilde G m^*}{4 \pi^3}
\int_{0}^\Lambda \frac{{\bf q}^2  \, dq}{E^*_{q}} \int_{-1}^{+1} dx
\biggl\lbrace  \mbox{P} \int_{0}^\infty dq_{0} \frac{\sum_{\phi}\mbox{Im}
\Pi^{(0)}_{v,\phi} (q_{0},\vert {\bf p} - {\bf q} \vert )}{q_{0}+p_{0}+E_{q}^*}
 \\ &&
+
\mbox{P} \int_{0}^\infty dq_{0} \frac{\sum_{\phi}\mbox{Im}
\Pi^{(0)}_{s} (q_{0},\vert {\bf p} - {\bf q} \vert )}{q_{0}-p_{0}+E_{q}^*}
\biggr\rbrace,\nonumber\label{eq:sigv2}
\end{eqnarray}
with
\begin{eqnarray}
\mbox{Im} \sum_{\phi}\Pi^{(0)}_{v,\phi} (q)& = &\mbox{Im}
\sum_{\phi}\left\lbrace -2i\tilde G \int \frac{d^4k}{(2\pi )^4} \mbox{tr}
\left[ \gamma^\phi g(k+q/2)\gamma_{\phi} g(k-q/2) \right] \right\rbrace
\\ &=&
+\frac{\tilde G n_{f}
n_{c}}{2 \pi^2} \int
\frac{d^3k}{E^*_{k}E^*_{k-q}} \left( E^*_{k}E^*_{k-q} + {\bf k} ({\bf
k}- {\bf q}) + 2 {m^*}^2 \right) \nonumber \\
&& \times \delta (\vert q_{0}\vert - E^*_{k} -
E^*_{k-q} ) \Theta (\Lambda - \vert {\bf k} \vert )
\Theta (\Lambda - \vert {\bf k } - {\bf q} \vert ) \; .
\end{eqnarray}
Since the imaginary part of the vector polarization propagator in the
preceeding equation is always positive, the real part of the
second-order scalar self-energy due to the vector
interaction has the same sign as the
analogous contribution due to the pseudoscalar interaction.

The results presented thus far can easily be extended to interaction
terms
which are isovector rather than the isoscalar in
eqs.(\ref{eq:scall}), (\ref{eq:pseudl}), and (\ref{eq:vecl}). If
\be
\tilde G \left( \bar\psi \hat O \psi \right)^2 \Rightarrow \tilde G \left(
\bar\psi \hat O \vec\tau \psi \right)^2,
\ee
then one obtains the corresponding second-order contribution to the scalar
part of the self-energy with an additional factor of 3, which
originates from the different flavor term in the interaction. If one
takes into account interaction terms which are products of vector
operators in SU(3) color space, then
\be
\tilde G \left( \bar\psi \hat O \psi \right)^2 \Rightarrow \tilde G \left(
\bar\psi \hat O \vec\lambda \psi \right)^2,
\ee
and the corresponding contributions to the self-energy should be multiplied
by a factor 32/9.

\section{Results and Discussion}

In order to display the basic features of the second-order
contributions to the scalar self-energy, a few plots obtained for the
scalar interaction (see eqs.(\ref{eq:scall}),
(\ref{eq:sigs1}), and (\ref{eq:sigs2})) are presented in figure 3.
In evaluating these contributions,
the Green's functions for the quarks have been approximated by  Green's
functions for Fermions with a constant effective mass $m^*$=313 MeV, and
the integrals have been regularized by a cut-off parameter $\Lambda$ =
653 MeV (see table II of ref.\cite{klev}).
This cut-off parameter will remain unchanged for all results presented
unless explicitly stated otherwise. The choice for the coupling
constant $\tilde G$ does not affect the qualitative features that
we want to discuss in the
first part of this section since all results for the second-order
contributions of the self-energy can be scaled by ${\tilde G}^2$. The
numerical values displayed in figure 3 were obtained by first employing $\tilde
G$=1.1088 $\times 10^{-5}$ MeV$^{-2}$ = $G_D$ and can be obtained
from eqs.(\ref{eq:gap}) and (\ref{eq:coupco}).
These self-energy values
were then scaled such that the result for the vector interaction,
eq.(\ref{eq:sigv2}), yielded a self-energy contribution of 1 MeV for
quarks with $\vert{\bf p}\vert = p_{0 }= 0$ (see figure 5 also).

Inspecting the results in figure 3 for Fermions with momentum
$\vert{\bf p}\vert$ = 0 as a function of the energy $p_{0}$,
one observes, centered about $p_{0}$ = 1500 MeV, a resonance structure
which is typical for a contribution to the self-energy of second- or
higher-order. This resonance structure simply reflects the fact that
diagrams, such as the one displayed in figure 1b), describe the modification
of the single-particle propagator due to the admixture of 2 particle -
1 hole (2p1h) configurations. This admixture provides a non-vanishing
imaginary part of the self-energy for energies $p_{0}$ which are
larger than the threshold of such 2p1h configurations. For the
quark self-energy with momentum ${\bf p}$ = 0, this threshold is
at $p_{0} = 3 m^*$ (= 939 MeV), while the largest energy leading to
a non-vanishing imaginary part
is determined by the cut-off parameter. The energy dependence of the
real part of the self-energy is determined by these characteristics of the
imaginary part because both are related to each other by a dispersion
relation. It is worth recalling that the self-energy is symmetric
with respect to
$p_{0}$ = 0: $\Sigma(-p_{0}) = \Sigma (p_{0})$.

For momenta $\vert{\bf p}\vert$ larger than zero, the energy dependence
is rather similar. In this case, however, one observes
that the absolute value of the real part of the self-energy is slightly
reduced as compared to the $\vert{\bf p}\vert$=0 case. This reduction
can be related to the fact that the phase space of 2p1h configurations
with momenta larger than zero is more strongly affected by the cut-off
than it is for $\vert{\bf p}\vert$=0.

Very similar features can be observed if a pure pseudoscalar
interaction (see eq.(\ref{eq:pseudl})) is considered. Results for the
real part of the scalar term of the self-energy are displayed in figure
4. Also, for these results, the same values of $m^*$,
$\Lambda$, and $\tilde G$ were employed as those discussed above for the
scalar interaction. Comparing the results of figures 3 and 4, one
notices that the real part of the self-energy for a
pseudoscalar interaction has the opposite sign as that of the
self-energy for a scalar interaction. This difference can be understood
by
comparing eqs.(\ref{eq:sigs1}) and (\ref{eq:pses1}). Furthermore,
one finds that, using the same coupling constant, the absolute value
of an arbitrary self-energy point
obtained from a pseudoscalar interaction is about twice as large as a
corresponding point obtained from the scalar interaction.
This difference
is reflected by the fact that the polarization propagators
$\Pi^{(0)}_{s}$ and $\Pi^{(0)}_{ps}$ differ (compare eqs.(\ref{eq:pols}) and
(\ref{eq:polps})). The dominance of the pseudoscalar contribution
with respect to the contribution obtained from a scalar interaction
is very important since
chiral symmetry always relates the coupling strengths for these two
interaction channels. Hence, we will always observe a partial
cancellation of the
second-order contributions to the self-energy obtained from the scalar and
pseudoscalar interactions;
however, their combined effect yields a positive contribution
to the real part of the self-energy for energies $p_{0}$ below the
threshold of 2p1h configurations.

Another positive contribution to the real part of the
scalar self-energy at
low energies is obtained for a pure vector
interaction of the form found in eq.(\ref{eq:vecl}). This fact is
illustrated in figure 5 by comparing self-energy plots obtained
from the three interactions utilized in this study. These self-energies
were calculated with
the same coupling constant $\tilde G$ and scaled as discussed earlier.
One finds that the contribution obtained from the vector interaction
is even larger in absolute value than
that obtained from the pseudoscalar interaction. Note, however, that
this fact does not lead to a model-independent conclusion concerning
the relative
importance of this contribution, since the coupling constant for a
vector interaction is independent of the interaction strength of the
scalar and pseudoscalar interactions: their ratio is not constrained by
chiral symmetry.

Again, with the same scaling procedure for the self-energies, it is apparent
that the dependence of the absolute value of the real parts
of the scalar self-energies
on the three-momentum $\vert{\bf p}\vert$ is very similar
for all three interactions. This fact is displayed in figure 6.

In the next step of our discussion, we turn to the various Lagrangians
discussed in section 2, which were denoted as models A, B, C, and D. The
coupling constant for each Lagrangian is now adjusted by
requesting that the sum of its Hartree-Fock and second-order contributions
leads to a
self-energy for $\vert{\bf p}\vert$ = 0 and $p_{0}$ = $m^*$ =
313 MeV which is identical to $p_{0}$:
\be
p_{0} = Re\Sigma_{s}^{\mbox{HF}} + \Sigma_{s}^{(2)} (p_{0},\vert{\bf p}\vert =
0).
\ee
As an example, we consider Lagrangian A (see eq.(\ref{eq:njla2})) for
which this equation takes the form:
\begin{eqnarray}
p_{0} & = & m_{0} + G_{A}\frac{13}{12}\frac{24}{(2\pi )^3} \int_{0}^\Lambda
d^3p \,
\frac{m^*}{\sqrt{{\bf p}^2 + {m^*}^2}} \nonumber\\ && + G_{A}^2 \Biggl[
\delta\tilde\Sigma_{s}^{(s)} (p_{0},{\bf 0}) \left\lbrace \left(
\frac{13}{12}\right)^2 + \left(\frac{1}{12}\right)^2 3 +
\left(\frac{1}{8}\right)^2 \frac{32}{9} +\left(\frac{1}{8}\right)^2
\frac{32}{3}\right\rbrace \nonumber\\ && +
\delta\tilde\Sigma_{s}^{(ps)} (p_{0},{\bf 0}) \left\lbrace \left(
\frac{1}{12}\right)^2 + \left(\frac{13}{12}\right)^2 3 +
\left(\frac{1}{8}\right)^2 \frac{32}{9} +\left(\frac{1}{8}\right)^2
\frac{32}{3}\right\rbrace \nonumber\\ && +
\delta\tilde\Sigma_{s}^{(v)} (p_{0},{\bf 0}) \left\lbrace \left(
\frac{1}{6}\right)^2 + \left(\frac{1}{4}\right)^2 \frac{32}{9}\right\rbrace
\Biggr] . \label{eq:detera}
\end{eqnarray}
In this equation, $\delta\tilde\Sigma_{s}^{(i)}$ represents the
second-order contribution to the scalar part of the self-energy
originating from an interaction of type $i$ ($i=s$, $ps$, or $v$),
calculated for a coupling constant $\tilde G=1$. The factors
multiplying these contributions refer to the actual value of the
coupling constant in this model (see eq.(\ref{eq:njla2})) multiplied by
the appropriate flavor-color factors, as discussed at the end
of section 3. The
Hartree-Fock term is calculated following the gap equation
(\ref{eq:gap}), and the current quark mass $m_{0}$ has been chosen to be
5 MeV. From eq.(\ref{eq:detera}), one can determine the new $G_{A}$ and
the coupling constants of the other models
in a corresponding way. This procedure
leads to the values which are quoted in the caption of table 1.

Table 1 lists the various contributions to the effective mass of
quarks with momentum ${\bf p}$ equal to zero if the coupling constant
is chosen according to procedure just outlined. By construction, the sum
of all contributions is equal to $m^*$ = 313 MeV. The importance of
individual components, however, depends to some extent on the Lagrangian
chosen
for the model. For example in our model A, the original NJL
Lagrangian of eq.(\ref{eq:njla1}), the effective mass is dominated by
the Hartree-Fock contribution. This can immediately be understood since
the coupling constant in front of the scalar-isoscalar-colorscalar
interaction term ( $(\bar\psi\psi)^2$ ) determines the importance of
the Hartree-Fock term. In eq.(\ref{eq:njla2}), which shows the
Lagrangian with inclusion of exchange terms, this constant is large
(13/12$G_{A}$) compared to the interaction strengths found in the other
interaction channels. Because of this large coupling constant in the
scalar and ``pionic'' channels, it is clear that, also, the
second-order contributions are dominated by the contributions of
these interaction channels (see first column of table 1). In this
model, however, the second-order contributions are small compared to
the leading HF term.

The situation is somewhat different for model B,
where it is assumed that the direct
interaction is a pure vector-isoscalar-colorscalar one
(eq.(\ref{eq:njlb1})). In this case the coupling constant defining the
strength in the scalar channel is rather weak (1/6 $G_{B}$) compared
to those in the other channels. But, also, in this case more than fifty
percent of the effective mass is due the Hartree-Fock contribution.

While for models A and D the coupling to the pseudoscalar, i.e.,
``pionic'' excitations is dominant, the largest contributions to the
second-order terms in models B and C originate from the coupling to
vector excitation modes. The second-order contributions describing
the coupling to the ``pionic'' excitations can be understood as a first
step towards an attempt to include the Fock contributions to the quark
self-energy due to pion exchange\cite{cao}. To include these pion
exchange contributions in a microscopic way, one would have to replace
the irreducible polarization propagator
$\Pi^{(0)}_{ps}$ in eq.(\ref{eq:pses1})
by the corresponding collective propagator. This would
lead to an imaginary part in the quark self-energy at lower energies
than the threshold discussed above and also shift the
characteristic energy dependence of the real part of the self-energy
displayed in figure 4 to lower energies. It should be kept in mind,
however, that the contribution to the self-energy originating from the
pionic excitations is only a part of the entire contribution.

An important difference between the HF contribution to the self-energy
and those originating in second-order is the fact that the
latter contributions vary with the energy and momentum of the quark
under consideration. In order to visualize these effects, we define an
effective mass $M_{\mbox{eff}}({\bf p})$
which characterizes the pole term in the Green's function for a quark
with momentum ${\bf p}$ by:
\begin{eqnarray}
p_{0}^2 & = & {\bf p}^2 + \left( m_{0} + Re \Sigma_{s}^{HF} +
\delta Re \Sigma_{s}^{(2)}(p_{0},{\bf p}) \right)^2 \\
& = & {\bf p}^2 + {M^2_{\mbox{eff}}({\bf p})} . \label{eq:meff}
\end{eqnarray}
Results for this effective mass are displayed in figure 7. Using the
coupling constants defined in table~1, all 4 models
must clearly provide an effective
mass of 313 MeV in the limit of vanishing three-momentum. Also, in each
model
under consideration, the effective mass decreases as a function of
increasing
three-momentum. This demonstrates that, for those combinations of
$p_{0}$ and ${\bf p}$ which solve eq.(\ref{eq:meff}), the
decrease of the second-order terms of the self-energy with momentum,
as displayed in figure 6, dominates the increase of these
contributions with increasing $p_{0}$, as displayed in figures 3-5. A
substantial enhancement is obtained only for values of $p_{0}$ which
are larger than those that solve eq.(\ref{eq:meff}). The momentum
dependence is of course larger for model B than for model A since the
contributions of second-order are more important in the former model.

Finally, we want to investigate the influence of the value chosen for
the cut-off parameter $\Lambda$. For that purpose, in table 2
we provide a comparison of the
various contributions to the self-energy obtained from a calculation for
$\Lambda$ = 800 MeV with the results discussed so far using $\Lambda$ =
653 MeV. The coupling constants for the various models with $\Lambda$ =
800 MeV are also adjusted to obtain an effective
mass $M_{\mbox{eff}}$ of 313 MeV in the limit of vanishing three-momentum.
An increase of the cut-off parameter yields a slight decrease in the
total
second-order contribution as compared to the leading HF term. All
other features discussed remain essentially the same.

\section{Conclusion}

The effects of second-order contributions to the real part of the
scalar self-energy of quarks are discussed in various models
inspired by the Nambu -- Jona-Lasinio (NJL) model. Depending on the explicit
form chosen for the Lagrangian, these second-order terms yield
contributions which range between 4 percent and 90 percent of the
leading Hartree-Fock contribution. The second-order contributions
depend on the energy and three-momentum of the quarks. This leads to a
momentum-dependent effective quark mass which decreases with increasing
momentum. The reduction can be as large as 20 percent for momenta near
the cut-off.

In calculating the second-order contributions, a partial cancellation is
observed between terms arising from the coupling to scalar excitations
on one hand and vector and pseudoscalar excitations on the other.
The relative importance of scalar and pseudoscalar contributions as
compared to the coupling to vector excitation modes is dictated by the
Lagrangian of the model considered. A restriction to the pionic modes,
i.e., to restrict the self-energy contributions to those arising from pion
exchange\cite{cao} is justified for certain choices of the Lagrangian
only.

The investigations presented here must be considered as a first step
towards a more detailed investigation of NJL inspired models beyond the
Hartree-Fock approximation. Presently, we have approximated the
calculation of the self-energy by assuming a Green's function for
quarks with a constant effective mass. In a consistent calculation, the
Green's function should be determined from the self-energy and account
for the momentum dependence of the effective mass\cite{rev,skour}.
Since, however, this momentum dependence is apparently moderate, we think
that the main features observed here should remain intact in such a
self-consistent calculation.

Furthermore, one should study the properties of mesons which are
obtained from a calculation of the various polarization propagators
which goes beyond the Hartree-Fock / RPA scheme that has been used until
now. As has already been discussed in the introduction (see figure
1), for such an investigation a consistent improvement of the quark
propagator and the residual interaction used to evaluate the
polarization propagators is required. In a next-step calculation,
such ``collective''
polarization propagators could be used to improve the calculation of
the self-energy.
Also, one may study other terms besides the scalar contribution to the
self-energy, investigate the effects of various cut-off procedures, and
explore the modifications obtained for non-vanishing baryon density and
temperature.

This work has partly been supported by the Deutsche
Forschungsgemeinschaft (DFG Fa 67/14-1) and the Graduiertenkolleg
``Struktur und Wechselwirkung von Hadronen und Kernen'' (DFG Mu
705/3-1).

\clearpage


\clearpage

\begin{table}
\caption{The various contributions to the real part of the scalar
self-energy. Listed are the contributions from the Hartree-Fock
term and the second-order contributions originating from the various
interaction terms. Results are presented for three-momentum
$\vert{\bf p}\vert$ = 0 and energy $p_{0}$ = $m^*$ = 313 MeV.
Different Lagrangians have
been considered for each of the models A, B, C, and D introduced in section 2.
In each case, the coupling constant $G$ has been adjusted in such a way
that the sum of all contributions listed yields $m^*$ = 313
MeV. The following coupling constants were obtained: $G_{A}$ = 4.38948
$\times 10^{-6}$ MeV$^{-2}$, $G_{B}$ = 1.53437
$\times 10^{-5}$ MeV$^{-2}$, $G_{C}$ = 4.80885
$\times 10^{-6}$ MeV$^{-2}$, and $G_{D}$ = 7.75355
$\times 10^{-6}$ MeV$^{-2}$. All entries are given in MeV.}
\begin{center}
\begin{tabular}{crrrr}
&&&&\\ \hline\hline
&&&&\\
\multicolumn{1}{c}{Term}&\multicolumn{1}{c}{Model A}&
\multicolumn{1}{c}{Model B}&\multicolumn{1}{c}{Model C}&
\multicolumn{1}{c}{Model D}\\
&&&&\\ \hline
&&&&\\
HF   &  297.204  &   159.830 &    267.159 &      215.376\\
&&&&\\
$(\bar\psi\psi)^2$ &
   -2.080 &    -0.602 &     -1.681 &        -1.092\\
$(\bar\psi\vec\tau\psi)^2$ &
-0.037 &    -1.805 &     -5.043 &        -3.277\\
$(\bar\psi\vec\lambda\psi)^2$ &
   -0.098 &    -4.813 &     -0.210 &       -16.523\\
$(\bar\psi\vec\tau\vec\lambda\psi)^2$ &
   -0.295 &   -14.439  &    -0.630  &       -0.410 \\
&&&&\\
$(\bar\psi i {\gamma}_{5}\psi)^2$ &
     0.021 &     1.031  &     2.880    &      1.872\\
$(\bar\psi i {\gamma}_{5}\vec\tau\psi)^2$ &
    10.694  &     3.093  &     8.641    &      5.616\\
$(\bar\psi i {\gamma}_{5}\vec\lambda\psi)^2$ &
      0.169   &   8.247   &    0.360    &      0.234\\
$(\bar\psi i {\gamma}_{5}\vec\tau\vec\lambda\psi)^2$ &
    0.506 &    24.742  &     1.080     &    89.491 \\
&&&&\\
$(\bar\psi  {\gamma}_{\mu}\psi)^2$ &
       0.213  &   109.945  &     1.818   &      18.901\\
$(\bar\psi  {\gamma}_{\mu}\vec\tau\psi)^2$ &
        0.0    &    1.952  &     5.453   &       0.0 \\
$(\bar\psi  {\gamma}_{\mu}\vec\lambda\psi)^2$ &
        1.704  &    5.205  &    27.492   &       2.363\\
$(\bar\psi  {\gamma}_{\mu}\vec\tau\vec\lambda\psi)^2$ &
       0.0    &   15.614  &     0.682   &       0.0 \\
&&&&\\ \hline\hline
\end{tabular}
\end{center}
\end{table}
\clearpage
\begin{table}[h]
\caption{Various contributions to the real part of the scalar
self-energy originating from Hartree-Fock and
second-order contributions
classified
according to
the Dirac structure only of the implemented interaction.
Results are compared for
quarks with momenta ${\bf p}$ = 0 choosing 2 different cut-off
parameters $\Lambda$. For further remarks, see the caption of table 1.}
\begin{center}
\begin{tabular}{crrrr}
&&&&\\ \hline\hline
&&&&\\
\multicolumn{1}{c}{Term}&\multicolumn{1}{c}{Model A}&
\multicolumn{1}{c}{Model B}&\multicolumn{1}{c}{Model C}&
\multicolumn{1}{c}{Model D}\\
&&&&\\ \hline
&&&&\\
\multicolumn{5}{c}{$\Lambda$ = 653 MeV} \\
&&&&\\
HF   &  297.204  &   159.830 &    267.159 &       215.376\\
scal.&  -2.511   &   -21.658 &    -7.564  &     -21.303   \\
ps.  &  11.390   &    37.113 &    12.961  &     92.663    \\
vec. &   1.917   &   132.715 &    35.444  &     21.264    \\
&&&&\\ \hline
&&&&\\
\multicolumn{5}{c}{$\Lambda$ = 800 MeV}   \\
&&&&\\
HF   &  297.764  &   163.752 &    269.608 &       218.292\\
scal.&  -2.845   &   -25.658 &    -8.694  &      -24.697  \\
ps.  &   11.264  &    38.381 &    13.005  &      93.783   \\
vec. &   1.816   &   131.524 &    34.080  &      20.623   \\
&&&&\\ \hline\hline
\end{tabular}
\end{center}
\end{table}
\clearpage
\section {Figure Captions}
\bigskip\bigskip\noindent{\bf Figure 1:}
{Diagrams displaying the Hartree-Fock self-energy (a) and the self-energy
of second-order in the interaction (b). The diagrams
displayed in figures c) through f) exhibit various contributions to the
polarization propagator as discussed in the text.}

\bigskip\bigskip\noindent{\bf Figure 2:}
{Diagrams representing the second-order contribution to the self-energy
for a Fermion with momentum $p$. The labels for the momenta of the
intermediate Fermion ($q$) and the polarization and boson propagators
($p-q$) are identical to those used in eq.(12).}

\bigskip\bigskip
\noindent{\bf Figure 3:}
{Real part of the scalar term of the self-energy for quarks with
momentum $p$, assuming a scalar interaction (eq.(13)). For
various values of $\vert{\bf p}\vert$, results are displayed as a
function of $p_{0}$. The self-energy has been calculated assuming a
Green's function for quarks characterized by a constant effective
mass $m^*$ = 313 MeV, a cut-off parameter $\Lambda$ = 653 MeV,
and a coupling constant $\tilde G$ = 1.1088 $\times 10^{-5}$
MeV$^{-2}$.}

\bigskip\bigskip\noindent{\bf Figure 4:}
{Real part of the scalar self-energy for quarks with various
momenta $\vert{\bf p}\vert$
assuming a pseudoscalar interaction (eq.(22)).
For further details, see the caption of figure 3.}

\bigskip\bigskip\noindent{\bf Figure 5:}
{Real part of the scalar self-energy for quarks with
momentum $\vert{\bf p}\vert$ =0, assuming various interaction terms.
For further details, see the caption of figure 3.}

\bigskip\bigskip\noindent{\bf Figure 6:}
{Real part of the scalar self-energy for quarks with
momentum $p$, assuming various interaction terms. Assuming $p_{0}$ = 1
GeV, results are presented as a function of $\vert{\bf p}\vert$.
For further details, see the caption of figure 3.}

\bigskip\bigskip\noindent{\bf Figure 7:}
{Effective mass $M_{\mbox{eff}}({\bf p})$ as defined in
eq.(34) for various momenta. Using the coupling constants provided
in the caption of table 1, the effective masses approach
the value 313 MeV in
the limit of zero three-momentum for all 4 models of the
Lagrangian.}

\end{document}